\documentclass{physeauth}
\usepackage{graphicx}
\usepackage{amsmath}
\usepackage{amssymb}

\makeatletter
    \renewcommand{\theequation}{%
    \thesection.\arabic{equation}}
    \@addtoreset{equation}{section}
  \makeatother

\begin{document}

\begin{frontmatter}

\title{Duality of quantum competing system} 

\author[mpi]{Morishige Yoneda},
\author[amst]{Masaaki Niwa} and
\author[amst]{Mitsuya Motohashi}

\address[mpi]{Denshi Gakuen Japan Electronics College, 1-25-4 Hyakunin-cho, Shinjuku-Ku,
Tokyo, Japan}
\address[amst]{Tokyo Denki University,2-2 Kanda-Nishiki-cho,Chiyoda-ku,Tokyo,Japan}

\thanks[thank1]{
Corresponding author. 
E-mail: yoneda@jec.ac.jp}

\begin{abstract}
  We have constructed a theory of dual canonical formalism to study the quantum competing systems. 
  In such a system, as the relationship between curWe have constructed a theory of dual canonical
   formalism to study the quantum competing systems. In such a system, as the relationship between
    current and voltage of each, we assumed the duality condition. We considered competing systems
     of two types. One type is composed of a sandwich structure with a SC (superconductor)/
     SI (superinsulator)/SC junction, and its dual junction consists of a sandwich structure formed
      by the SI/SC/SI junction. The other type of system consists of a sandwich structure formed by
       the SC/FM (ferromagnet)/SC junction, and its dual junction consists of a sandwich structure
        formed by the FM/SC/FM junction (spin Josephson junctions).We derived the relationship
         between the phase and the number of particles in a dual system of each other.  
         As an application of the dual competitive systems, we introduce a quantum spin transistor.
\end{abstract}

\end{frontmatter}


\section{Introduction}
In superconducting systems, there is a Josephson junction device known as a quantum effect devices[1,2]
 that operates by means of quantum flux tunneling. The mesoscopic Josephson junction is a quantum effect
  device that operates by using single Cooper pair tunneling created by a Coulomb blockade[3]. 
  The junction by the superconductor as the condensation of Cooper pairs and the superinsulator as 
  the condensation of the quantum flux (vortex), i.e., the junction of superconductor and superinsulator
   that are dual to each other[4,5], forms a competing system. On the other hand, like magnetic duality,
    spontaneous magnetization and the domain wall are known to have a dual relationship. The junction 
    is formed by the ferromagnet as the condensation of the spin magnetization and by the superconductor
     as the condensation of the domain wall; therefore, the junction[6] of the ferromagnet and the perfect
      diamagnetism (superconductor), where they are dual to each other, forms a competing system.
        As described above, we considered competing quantum systems of two types, and as an application
         of these systems, we propose a quantum spin transistor. Our main objective is to build a theory
          of quantum devices, in which the freedom of a dual particle plays an important role, and we would
           also like to build upon the duality theory of competitive systems. This paper is composed as 
           follows. In the next section, as the duality system of the charge and magnetic flux,
            we introduce the dual canonical formalism[7] between the SC/SI/SC junction and 
            the SI/SC/SI junction, and by imposing the dual condition to between these, we derived
             the quantum resistance. In sec.3, we derive the partition functions by means of the path 
             integral for a quantum single Josephson junction and its dual model, respectively. 
             Insec.4, as the duality of the spin and magnetic domain wall, we study the FM/SC/FM[6]
              junction and its dual model as an analogy with the Josephson junctions. In Sec.6, 
              as an application of the junction mentioned in the previous section, we introduce a quantum
               spin transistor. In the last section, we present a summary and conclusions.
\section{Dual canonical formalism between SC/SI/SC junction and SI/SC/SI junction}
\label{sec:Ratchets}
Duality has been known to be a powerful tool in various physical systems such as statistical 
mechanics[8] and field theory[9,10]. Furthermore, it has been also recognized in several studies 
of Josephson junction systems[11,12]. In this section, we introduce the dual canonical formalism[7] 
between a SC/SI/SC junction and a SI/SC/SI junction. First, we propose a small SC/SI/SC[7] 
junction and its equivalent circuit, which are shown in Fig.1. 
\begin{figure}[tb] 
\begin{center} 
   \includegraphics[angle=0,width=.38\textwidth]{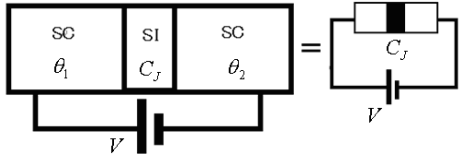}
    \caption{ Schematic of SC/SI/SC junction and its equivalent circuit.} 
   \label{fig:r} 
  \end{center} 
\end{figure} 
It consists of the sandwich structure of 
the SC/SI/SC. We introduce the order field, which is given by  $\psi\!\!\equiv\!\!\sqrt{N}\!\exp\!\left(i\theta \right)$.
In superconducting systems, for example, $\,N\!\!\equiv\!\!{{N}_{\!1}}\!-\!{{N}_{\!2}}$ represents the relative number
 operator of a Cooper pair, and $\theta\!\!\equiv\!\!{{\theta }_{\!1}}\!-\!{{\theta }_{\!2}}$ 
 represents the relative phase of Cooper pairs. These commutation relation is given 
 by $[\ \theta ,N ]\!=\!i$. The junction is characterized by 
 its capacitance ${{C}_{\!J}}$ and the Josephson energy $\,{{E}_{\!J}}\!\!=\!\!{\hbar {{I}_{\!c}}}/{2e}$, 
 where ${{I}_{\!c}}$ is the critical current. Now, when this junction system is considered as 
  the Josephson junctions, it can be described by the Hamiltonian of a quantum single Josephson
   junction as follows:
\begin{eqnarray}
        H=4{{N}^{2}}{{E}_{\!c}}+{{E}_{\!J}}\left(1-\cos \theta  \right).        
\end{eqnarray}
The first term describes the Coulomb energy of the Josephson junction, where  $\,N$ 
is the number operator of the Cooper pair, and $\,{{E}_{c}}$$\equiv$ $\,{{{e}^{2}}}\!/{2C}$ 
is the charging energy per single-charge. The second term describes the Josephson coupling
 energy. From eq.(2.1), the Josephson equations are given by
$\,V\!\!\equiv\!\!\left( {\hbar }/{2e}\right){\partial \theta }/{\partial t}\!=\!{4N}{{E}_{c}}/{e}$,
$\,I\!\equiv\!\left( 2e \right){\partial N}/{\partial t}\!=\!-{{I}_{c}}\sin \theta$.
where $\,V$ and $\,I$ are the voltage and current of the Cooper pair, respectively.  
Next, we describe the theoretical model and basic equations for a SI/SC/SI junction.
 In such a system, we propose a small SI/SC/SI junction and its equivalent circuit,
  as shown in Fig.2. It consists of the sandwich structure of the SI/SC/SI. 
  We introduce a dual particle field[4,13] 
 $\tilde{\psi }\!\!\equiv\!\!\sqrt{\!{\tilde{N}}}\exp( i\tilde{\theta } )$.
  In superconducting systems, for example, 
 $\tilde{N}\!\equiv\!{{\tilde{N}}_{1}}\!-\!{{\tilde{N}}_{2}}$
 represents the relative number operator of a vortex, and
  $\tilde{\theta }\equiv {{\tilde{\theta }}_{1}}-{{\tilde{\theta }}_{2}}$ 
  represents the relative phase of a vortex. These commutation
   relation is given by $[\ \tilde{\theta },\tilde{N}]=i$. 
  The dual Hamiltonian of eq.(2.1) is given by
  \begin{eqnarray}
\tilde{H}={{\tilde{N}}^{2}}{{E}_{v}}+\frac{2{{E}_{c}}}{{{\pi }^{2}}}\left( 1-\cos \tilde{\theta } \right),
  \end{eqnarray}
Here, the first term describes the vortex energy of the dual Josephson junction, where 
${{E}_{v}}\!\equiv\!2{{\pi }^{2}}{{E}_{J}}$ $\!=\!\Phi _{0}^{2}/2{{L}_{c}}$
is vortex energy per single-vortex and ${{L}_{c}}\!\!\equiv\!\!{{\Phi }_{0}}/2\pi {{I}_{c}}$ 
is the critical inductance. The second term describes the dual Josephson coupling energy. 
From eq.(2.2) dual Josephson equations are given by
$\tilde{V}\!\!\equiv\!\!\left( {\hbar }/{{{\Phi }_{0}}}\right)\!{\partial \tilde{\theta }}/{\partial t}\!=\!2\pi {{I}_{c}}\tilde{N},$ 
$\tilde{I}\!\!\equiv\!\! -{{\Phi }_{0}}{\partial \tilde{N}}/{\partial t}\!=\!{2{{E}_{c}}}\!\sin \tilde{\theta }/\!{\pi e},$ 
\begin{figure}[tb] 
\begin{center} 
    \includegraphics[angle=0,width=.39\textwidth]{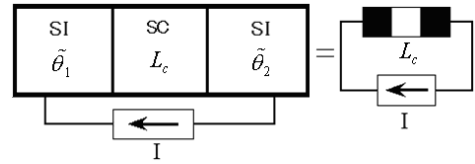}
    \caption{ Schematic of SI/SC/SI junction and its equivalent circuit.} 
   \label{fig:r} 
  \end{center} 
\end{figure} 
where $\tilde{V}$ and $\tilde{I}$ are voltage of vortex and current of vortex respectively.
Here, we assume the following duality conditions,
\begin{eqnarray}
  \hspace{0.5em}V\equiv \tilde{I},
  \hspace{0.5em}I\equiv \tilde{V}. 
\end{eqnarray}
By imposing these duality conditions, we derived the next two types of relationships.
 One is the relationship between the phase of the Cooper pair $\theta $ and 
 the vortex number $\tilde{N}$, the other type is the relationship between 
 the phase of the vortex field  $\tilde{\theta }$ 
and the Cooper pair number $N$ which, is given as follows:
 \begin{eqnarray}
\tilde{N}={-\sin \theta }/{2\pi }\;,
N={\sin \tilde{\theta }}/{2\pi }\;, 
\end{eqnarray}
From the Josephson equation, we derived the resistance given by
 \begin{eqnarray}
R=\frac{-\hbar }{{{(2e)}^{\!2}}}\frac{8N}{\sin \theta }\frac{{{E}_{\!c}}}{{{E}_{\!J}}}
=\frac{2{{R}_{Q}}}{{{\pi }^{\!2}}}\frac{N}{{\tilde{N}}}\frac{{{E}_{\!c}}}{{{E}_{\!J}}},
\end{eqnarray}
where ${{R}_{\!Q}}\!\equiv\! h/{{(2e)}^{2}}$ is the quantum resistance [14,15]. 
In the last equality in eq.(2.5) , we used the relationship of eq.(2.4). 
In the same manner, from the dual Josephson equations, we derived the conductance
 $\tilde{R}\!=\!{{R}^{-1}}$. In the case of the condition of 
 $\tilde{N}\!\gg\!N$ or ${{E}_{c}}\!\gg\!{{E}_{J}}$,
  , which is the state in a insulator. In particular, in this extreme case $R\!\to\!\infty$,
   which is the state in a superinsulator. In the reverse case, in the condition of $\tilde{N}\!\ll\!N$
    or ${{E}_{J}}\!\gg\!{{E}_{c}}$, which is the state in a conductor . In particular,
     in this extreme case  $R\!\to\!0$,which is the state in a superconductor. 
     As a special case of these conditions, in the case of
      $\tilde{N}\!=\!N$ and ${{E}_{c}}\!=\!{{{E}_{J}}{{\pi }^{2}}}/{2}\;$, in which the resistance $R$ 
      is equal to the quantum resistance ${{R}_{Q}}$, which is the state in a self dual. In the above discussion,
       we have shown that we can more clearly define the presence of quantum resistance and a quantum critical 
       point by using a dual canonical formalism.
       
\section{The partition function by path integral of a quantum single Josephson junction and its dual model}
\label{sec:RW}
In this section, using the Hamiltonian of the quantum single Josephson junction shown in Sec.2, 
we derive the partition function of a quantum single Josephson junction. From the Hamiltonian 
in eq.(2.1), the partition function $Z\left( \beta  \right)$ of a quantum single Josephson 
junction with imaginary time $\tau$ is as follows:
\begin{eqnarray}
Z\!=\!\!\!\!\int\!\!\!\!{D}\!\hbar\!N\!\!\!\!\int\!\!\!\!{D}\!\theta\!\exp\!\!\!\int_{0}^{\hbar \beta}
\!\!\!{d\tau }\!\!\left\{\!\!i\hbar \frac{\partial \theta}{\partial \tau }\!N
\!-\!4{{E}_{c}}{{N}^{2}}\!\!-\!{{E}_{\!J}}\!\left(\!1\!-\!\cos \theta\right) \right\},\nonumber\\
\end{eqnarray}
As a result of integration by $\theta$, the partition function is as follows:
$Z\!\!\!\!=\!\!\!\!\int\!\!{D}\hbar \!N\!\exp\!\!\int_{0}^{\hbar \beta }
\!\!{d\tau }\!\!\left\{\!-4{{E}_{\!c}}{{N}^{2}}
\!\!+\!\ln\!\left[ {I}_{\alpha}\!\!\left( {{E}_{\!J}} \right) \right] \right\}.$ 
Here ${I}_{\!\alpha }\!\!\left( {{E}_{\!J}} \right)$ 
are the modified Bessel functions of the $\alpha$-th order, and $\alpha$ are defined as $\alpha\!\!\equiv\!\!-\hbar \partial N/\partial \tau $. 
We investigated the partition function using the Villain approximation[16] as follows:
${I}_{\!\alpha}\!\left( {E}_{\!J} \right)\!\!\cong\!\!{I}_{0}\!\left( {{E}_{\!J}}
 \right)\exp\!\left[ {-{{\!\alpha }^{2}}}\!/{2{{\left( {E}_{\!J} \right)}_{v}}}\right],$
where ${{\left( {E}_{J} \right)}_{v}}$ are Villain's constants as defined by  
${{\!\left( {E}_{\!J} \right)}_{\!v}}\!\!\equiv\!\!{-1}/{2\ln\!\left[ {{{I}_{\!\alpha }}
\!\left( {E}_{\!J} \right)}/{{{I}_{0}}\!\left( {E}_{\!J} \right)}\;\!\right]},$ 
where ${I}_{0}\!\left( {E}_{\!J} \right)$ 
are modified Bessel functions of the 0-th order. By using the Villain approximation, we can integrate out
$N$ in the partition function. The partition function then becomes as follows:
\begin{eqnarray}
Z={{{\left[ {{I}_{0}}\left( {{E}_{J}} \right) \right]}^{{\hbar \beta }
/{\varepsilon }\;}}}/{\sinh \left( \beta \sqrt{2{{\left( {{E}_{J}} \right)}_{v}}{{E}_{c}}} \right)}\;,   
\end{eqnarray}
where $\varepsilon$ is a slice unit of imaginary time and $M\equiv {\hbar \beta }/{\varepsilon }\;$
 is the total number of slice units. In addition, we can directly integrate out $N$ in the eq.(3.1). 
 Then, the Lagrangian $L( \dot{\theta },\theta )$ is expressed as a function of only theta as follows:
 \begin{eqnarray}
L\!=\!\frac{1}{2}\ln \left( \frac{{{\hbar }^{2}}\pi }{4{{E}_{c}}} \right)
\!+\!\frac{-{{\hbar }^{2}}}{16{{E}_{c}}}{{\left( {{\partial }_{\tau }}\theta  \right)}^{2}}
\!-\!{{E}_{J}}\left(\!1\!-\!\cos \theta  \right),
\end{eqnarray}                                     
On the other hand, we applied the variables transformation of eq.(2.4) to the partition function of eq.(3.1),
 and we integrated out $\theta$. Then, the Lagrangian $L( \dot{\tilde{\theta }},\tilde{\theta })$ 
 is expressed as a function of only theta tilde as follows:
\begin{eqnarray}
L\!=\!\ln\!\left( \frac{\hbar }{2\pi }\cos \tilde{\theta } \right)
\!+\!\ln \left[ {{I}_{\chi}}\left( {{E}_{J}} \right) \right]\!-\!\frac{4{{E}_{c}}}{{{\left( 2\pi  \right)}^{2}}}{{\sin }^{2}}\tilde{\theta },
\end{eqnarray}
where $\chi$ are defined as $2\pi \chi\!\!\equiv\!\!-\hbar \cos \tilde{\theta }{\partial \tilde{\theta }}/{\partial \tau }.\;$
By compare with eq.(3.3) and eq.(3.4), we find that these Lagrangians are a mutually dual representation.
Next, using the dual Hamiltonian in eq.(2.2), we derive the partition function 
$\tilde{Z}\left( \beta  \right)$ of a single dual quantum Josephson junction as follows:
\begin{eqnarray}
\tilde{Z}\!\!=\!\!\!\int\!\!\!{D}\!\hbar\tilde{N}\!\!\!\!\int\!\!\!{D}\tilde{\theta }
\!\exp\!\!\!\int_{0}^{\hbar \beta }\!\!\!\!\!{d\tau }\!\!\left\{\!i\hbar \frac{\partial \tilde{\theta }}{\partial \tau }\!\tilde{N}
\!\!-\!\!{{E}_{v}}\!{{\tilde{N}}^{2}}\!\!-\!\!\frac{2{{E}_{c}}}{{{\pi }^{2}}}\!\left(\!1\!-\!\cos\!\tilde{\theta } \right)\!\!\right\},\nonumber\\
\end{eqnarray}
As a result of integration by $\tilde{\theta }$, the partition function is as follows:
$\tilde{Z}\!\!=\!\!\!\int\!\!{D}\hbar \tilde{N}\exp\!\!\int_{0}^{\hbar \beta }\!\!\!{d\tau }\!\!
\left\{\!-{{E}_{v}}\!{{\tilde{N}}^{2}}\!\!\!+\ln\!\! \left[ {{I}_{\tilde{\alpha }\!}}
\!\left( {2{{E}_{c}}}\!/\!{{{\pi }^{2}}}\; \!\!\right) \right] \right\},$
where $\tilde{\alpha }$ are integer fields as $\tilde{\alpha }\equiv -\hbar \partial \tilde{N}/\partial \tau $.
 Using the same procedure as in the previous section, we derived the dual partition function as follows:
\begin{eqnarray}
 \tilde{Z}\!\left( \beta  \right)\!\!=\!\!{{{\left[ {{I}_{0}}\!\!\left(\!{2{{E}_{c}}}/{{{\pi }^{2}}\!\!}\; \!\right) \right]}
 ^{\!{\hbar\!\beta }\!/{\varepsilon }\;}}}\!\!\!\!/{\sinh\!\!\left(\!\!\beta\!\sqrt{{{\!\left(\!{2{{E}_{c}}}/{{{\pi }^{2}}}\;\!\!\right)}_{v}}{{\pi }^{2}}\!{{E}_{J}}} \right)}\;\!\!. 
\end{eqnarray} 
If we compare eq.(3.2) and eq.(3.6), they are seen to be equal under the self dual conditions of  
 ${{E}_{c}}={{{E}_{\!J}}{{\pi }^{2}}}/{2}\;$. As with the derivation of eq.(3.3), we can integrate out 
 $\tilde{N}$ in eq.(4.1). Then the Lagrangian $\tilde{L}(\dot{\tilde{\theta }},\tilde{\theta })$ 
 is expressed as a function of only theta tilde as follows:
\begin{eqnarray}
\tilde{L}\!=\!\frac{1}{2}\ln\!\left(\! \frac{\pi {{\hbar }^{2}}}{2{{\pi }^{2}}{{E}_{\!J}}} \!\right)
\!\!+\!\frac{-{{\hbar }^{2}}}{8{{\pi }^{2}}\!{{E}_{\!J}}}{{\left( \!{{\partial }_{\!\tau }}\tilde{\theta }\! \right)}^{2}}
\!\!\!-\!\!\frac{2{{E}_{c}}}{{{\pi }^{2}}}\!\left( \!1\!-\!\cos \tilde{\theta }\right).
 \end{eqnarray} 
On the other hand, as with the derivation of eq.(3.4), we applied the variables transformation of eq.(2.4) 
to the partition function of eq.(3.5), and we integrated out $\tilde{\theta }$. Then the Lagrangian 
$\tilde{L}( \dot{\theta },\theta)$ is expressed as a function of only theta tilde as follows:
\begin{eqnarray}
\tilde{L}\!=\!\ln\!\left( \frac{\!-\hbar\! }{2\pi }\!\cos\!\theta  \right)
\!\!+\!\ln \left[ {{I}_{\tilde{\chi }}}\left( {2{{E}_{c}}}/{{{\pi }^{2}}}\; \!\!\right) \!\right]
\!-\!\frac{{{E}_{\!J}}}{2}{{\sin }^{2}}\theta,
\end{eqnarray}
where $\tilde{\chi }$ are defined as $2\pi \tilde{\chi }\equiv \hbar \cos \theta {\partial \theta }/{\partial \tau }\;$. 
By comparing eq.(3.7) and eq.(3.8), we find that these Lagrangians are a mutually dual representation. In addition, 
at the limit of large ${{E}_{J}}$ in eq.(3.4), if these satisfy the Gaussian approximation of $\tilde{\theta }$, 
eq.(3.4) and eq.(3.7) are the same in the above conditions. Similarly, at the limit of large ${{E}_{c}}$ 
in eq.(3.8), if these satisfy the Gaussian approximation of $\theta $, 
eq.(3.8) and eq.(3.3) are the same in the above conditions.

\section{The FM/SC/FM junction and its dual model as an analogy with the Josephson junctions}
\begin{figure}[tb] 
\begin{center} 
    \includegraphics[angle=0,width=.4\textwidth]{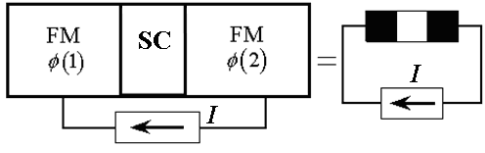}
    \caption{ Schematic of FM/SC/FM junction and its equivalent circuit.} 
   \label{fig:r} 
  \end{center} 
\end{figure} 
Thus far, we have dealt with models of a quantum single Josephson junction and its dual model 
to study the dual canonical formalism. In this chapter, as the duality of the spin and magnetic 
domain wall, we propose a single quantum spin device that operates using single quantum spin tunneling.
 The single spin transistor consists of the sandwich structure of the FM/SC/FM junction. In an analogy 
 with a Josephson junction, the FM/SC/FM junction can be thought of as a ferromagnetic junction with 
 a superconducting thin film barrier.
In this case, the superconducting thin film functions as a spin capacitor. 
As a model for such ferromagnetic junction systems, first, we consider the Hamiltonian
 of the Heisenberg XXZ spin models, as follows:
${{H}_{\!F\!M\!}}\!\!=\!\!\sum\limits_{\left\langle i,j \right\rangle }\!\left[\!{-{{J}_{\!xy}}\!\left(S_{i}^{x}S_{j}^{x}
\!+\!S_{i}^{y}S_{j}^{y} \right)}\!+\!{{J}_{\!z}}S_{i}^{z}S_{j}^{z} \right],$
Here, the first term describes the Ising spin energy and the second term describes the XY spin energy.
 We can rewrite the Hamiltonian with the introduction of a spin coherent state, and thus derive the Hamiltonian
  of the quantum single spin junction as follows:
 \begin{eqnarray}
{{H}_{\!F\!M\!}}={{E}_{sc}}{{N}_{\!X\!Y\!}}^{2}+{{E}_{xy}}\left( 1-\cos \phi  \right),
\end{eqnarray}
where ${{E}_{sc}}\!\equiv\!{{(S_{z}^{0})}^{2}}\!/2{{C}_{s}}$ is the spin charging energy per single spin,
 $S_{z}^{0}\!\equiv\!{\hbar }/{2}\;$ is the spin quantum unit, ${{N}_{\!X\!Y\!}}\!\equiv\!S_{^{z}}^{{}}/S_{^{z}}^{0}$
  is the relative number operator of XY ferromagnet quasi particles, and ${{C}_{s}}\!\equiv\!1/2{{J}_{z}}$
   is the spin capacitance. The ${{J}_{xy}}\!\!\propto\!\!{{E}_{\!xy}}\!\!\equiv\!\!{\hbar I_{s}^{c}}/{S_{\!z}^{0}}\;$
   in the second term of eq.(4.1) is the junction energy of a single spin junction, where 
   $I_{c}^{s}\!\equiv\!{S_{\!z}^{0}\!{{E}_{\!xy}}}/{\hbar }\;$ 
   is the critical spin current, and $\phi$ represents the relative phase of these quasi XY spin particles.
    Eq.(4.1) can be interpreted as the energy band of quasi spin particles in a periodic potential
    ${{E}_{xy}}\!\left(\!1-\cos \phi  \right)$. In particular, in the case of ${{E}_{sc}}\!\gg\!{{E}_{xy}}$, 
    eq.(4.1) is similar to the Bloch wave oscillations in a small Josephson junction. 
     Figure.4 shows the energy band that takes the energy on the vertical axis and spins a magnetic moment 
     on the horizontal axis. Using the Hamiltonian in eq.(4.1), spin Josephson equations
      are given by
\begin{figure}[tb] 
\begin{center} 
   \includegraphics[angle=0,width=.44\textwidth]{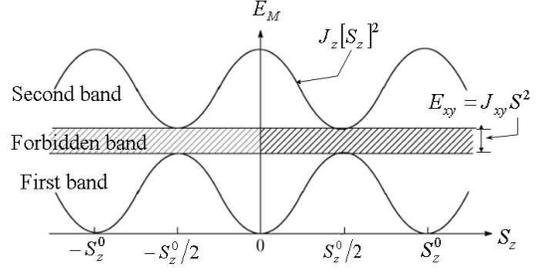}
    \caption{ The energy band to the ferromagnetic system.} 
   \label{fig:r} 
  \end{center} 
\end{figure} 
${{V}_{s}}\equiv ({\hbar }/{S_{z}^{0}}){\partial \phi }/{\partial t}={2{{N}_{\!X\!Y\!}}}{{E}_{sc}}/{S_{z}^{0}},$            
${{I}_{s}}\equiv {\partial {{S}_{z}}}/{\partial t}=-I_{_{z}}^{c}\sin \phi,$ 
where ${{V}_{s}}$ and ${{I}_{s}}$ are the spin voltage and spin current of the XY ferromagnetic spin, respectively.
 The approximate commutation relations between  ${{N}_{XY}}$ and $\phi$ are given by 
  $[\ \phi ,{{N}_{\!\!X\!Y}}]\approx i$. 
  The dual Hamiltonian of eq.(4.1) is given by:
\begin{eqnarray}  
{{H}_{dw}}={{E}_{dw}}{{N}_{\!D\!W\!}}^{2}+{{\tilde{E}}_{xy}}\left( 1-\cos \tilde{\phi } \right).
\end{eqnarray} 
In the first term, ${{E}_{\!dw}}\!\!\equiv\!\!{\Phi _{dw}^{2}}/{2L_{s}^{c}}\;$ describes the domain wall energy,
 where $L_{s}^{c}\!\equiv\!{4}/{{{E}_{xy}}}\;$is the spin inductance, $\Phi _{dw}^{{}}$ is the domain wall,
  and ${{N}_{\!D\!W}}\!\!\equiv\!\!\Phi _{\!dw}^{{}}\!/\Phi _{\!dw}^{0}$ is the number operator of the domain wall, 
  where $\Phi _{dw}^{0}\!\equiv\!4\pi$ is the domain wall quanta. In the second term of eq.(4.2),
     ${{\tilde{E}}_{xy}}\!\!\equiv\!\!{{{E}_{\!sc}}}/{2{{\pi }^{2}}}\;$is
      the junction energy of a single domain wall junction, and ${{N}_{\!D\!W}}$ and $\tilde{\phi }$ 
     represent the relative phase of the domain wall fields. Using the Hamiltonian in eq.(4.2), 
     dual spin Josephson equations are given by      
${{\tilde{V}}_{\!s}}\!\!\equiv\!\!({\hbar }/{\Phi _{dw}^{0}}){\partial \tilde{\phi }}/{\partial t}\!\!=\!\!I_{s}^{c}2\pi {{N}_{\!\!D\!W\!}},$
${{\tilde{I}}_{\!s}}\!\equiv\!-\Phi _{\!dw}^{0}{\partial {{N}_{\!D\!W\!}}}\!/{\partial t}\!\!=\!\!{{{E}_{sc}}}\sin \tilde{\phi }/{\pi S_{z}^{0}}.$
where ${{\tilde{V}}_{s}}$ and ${{\tilde{I}}_{s}}$ are the dual spin voltage and the dual spin current of 
the dual XY ferromagnetic spin, respectively. The approximate commutation relations between ${{N}_{\!D\!W}}$ 
and  $\tilde{\phi }$ are given by $[ \tilde{\phi },{{N}_{\!\!D\!W}}]\!\approx\! i$.
  By imposing similar duality conditions, ${{V}_{s}}\equiv {{\tilde{I}}_{s}}$,
  ${{I}_{s}}\equiv {{\tilde{V}}_{s}}$ to eq.(2.4), we derived the next two types of relationship,
\begin{eqnarray}
{{N}_{\!D\!W}}={-\sin \phi }/{2\pi }\;,                  
{{N}_{\!X\!Y}}={\sin \tilde{\phi }}/{2\pi }\;.                   
\end{eqnarray}
From the spin Josephson equations, we derived the spin resistance
${{R}_{s}}\!\!\equiv\!\!{{V}_{s}}/{{I}_{s}}\!\!=\!\!
{R_{Q}^{s}}({{N}_{\!X\!Y}}/{{N}_{\!D\!W}})({{E}_{sc}}/{2{{\pi }^{2}}{E}_{xy}}),$
where $R_{Q}^{s}\equiv h/{{(S_{z}^{0})}^{2}}$ is the quantum spin resistance.
In the same manner, from the dual spin Josephson equation, we derived the spin conductance
 ${{\tilde{R}}_{s}}\!\!\equiv\!\!{{{{\tilde{V}}}_{s}}}/{{{\tilde{I}}}_{s}}\!\!=\!\!{{R}_{s}}^{\!\!-1}.$ 
 In the case of a condition of
${{N}_{\!X\!Y}}\!\!\gg\!\!{{N}_{\!D\!W}}$ or ${{E}_{sc}}\!\!\gg\!\!{{E}_{xy}}$, 
 which is the state in a spin insulator. In particular, in the extreme case of ${{R}_{s}}\!\!\to\!\!\infty$, 
 which is the state in a super spin insulator. In the reverse case, in the condition of 
 ${{N}_{\!X\!Y}}\!\!\ll\!\!{{N}_{\!D\!W}}$ or  ${{E}_{xy}}\!\!\gg\!\!{{E}_{sc}}$, 
 which is the state in a spin conductor. In particular, in the extreme case of ${{R}_{s}}\!\!\to\!\!0$,
 the state in a super spin conductor. As a special case of these conditions, in the case of 
 ${{N}_{\!X\!Y}}\!\!=\!\!{{N}_{\!D\!W}}$ and ${{E}_{sc}}\!\!=\!\!2{{\pi }^{2}}\!{{E}_{xy}}$, 
 the spin resistance ${{R}_{s}}$ is equal to the quantum spin resistance $R_{Q}^{s}$, 
 which is the state in a self dual.

\section{The quantum spin transistor}
The single electron transistor[17] is a quantum effect device that operates by using the single electron 
tunneling created by a Coulomb blockade. Analogously, we propose a single quantum spin transistor that 
operates by using the single spin tunneling created by a spin blockade. The single quantum spin transistor 
consists of the FM/SC/FM junction of the sandwich structure. In this section, we consider the mechanisms 
that lie behind the single quantum spin tunneling, and advance a theoretical analysis to control this device. 
Now, ${{V}_{\!s}}$ is the dimension of the frequency ${{V}_{\!s}}\!\!=\!\!{2\pi I}\!/{e}\!\!=\!\!2\pi {{f}_{\!s}}$, where ${{f}_{s}}$ 
is the frequency of the single electron tunneling oscillations and ${{I}_{\!s}}$ 
  is the dimension of the energy ${{I}_{s}}\!\!=\!\!{e\!V}\!/\!{2\pi }\;$. Here, we used the following assumptions: 
  $\phi \!\propto\! \tilde{\theta }$ and $\tilde{\phi }\!\propto\! -\theta$. 
  As shown in Fig.5, as an application of the FM/SC/FM junction, we have devised a spin transistor, 
  where inductance $L$ of the FM/SC/FM junction is defined by $L\equiv{2{{\pi }^{2}}}/{{{J}_{z}}{{e}^{2}}};$ 
  ${{L}_{1}}$ and ${{L}_{2}}$ are the inductance of junction1 and junction2, respectively;
   ${{I}_{g}}$ is the current of the gate current source and ${{L}_{g}}$ 
   is the inductance of the gate current source. In each of junction1 and junction2, the forbidden condition
    of one quantum spin tunneling is given by the following respective equations:  
$I\!\!\!=\!\!\!\left\{ \pm \pi S_{z}^{0}\!+\!2\pi S_{z}^{0}\left( \!{{N}_{1}}\!-\!{{N}_{2}}\! \right)
\!+\!e{{L}_{g}}\!{{I}_{g}}\! \right\}\!/\!{e\left( \!{{L}_{2}}\!+\!{{L}_{g}}\! \right)},$        
$I\!\!=\!\!\left\{ \pm \pi S_{z}^{0}\!-\!2\pi S_{z}^{0}(\!{{N}_{1}}
\!-\!{{N}_{2}}\!)\!-\!e{{L}_{g}}\!{{I}_{g}}\!\right\}\!/\!{e{L}_{1}}.$ 
For the above conditions, Fig.6 shows the operating characteristics of the FM/SC/FM junction. 
Analogous to the Coulomb diamond in the Coulomb blockade, in the area inside the diamond, 
the tunneling of spin is blocked. This means that there is a real spin blockade without electron tunneling. 
\begin{figure}[tb] 
\begin{center} 
   \includegraphics[angle=0,width=.36\textwidth]{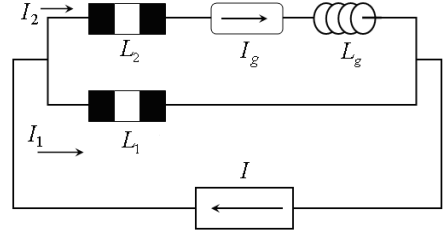}
    \caption{ Schematic of quantum spin transistor} 
   \label{fig:r} 
  \end{center} 
\end{figure} 
\begin{figure}[tb] 
\begin{center}
   \includegraphics[angle=0,width=.45\textwidth]{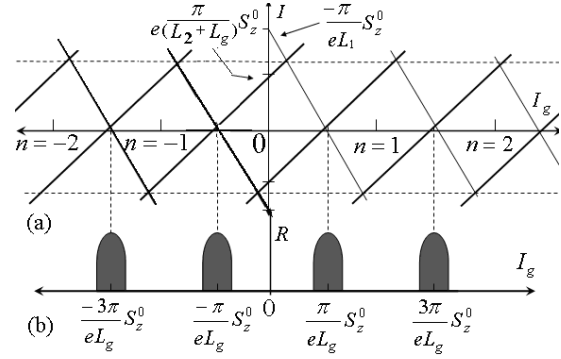}
\caption{ (a) Current ${I}$ as a function of ${{I}_{g}}$, the area inside the diamond,
 tunneling of spin is blocked. (b) Resistance ${R}$ as a function of ${{I}_{g}}$ for the quantum spin transistor.} 
   \label{fig:r} 
  \end{center} 
\end{figure} 
\section{Summary and Conclusion}
The results shared in this paper can be described as follows. As our first result, 
we showed examples of a competing system with each other of two types. In one of them,
 as the electrical duality, we introduce a duality between the superconductor and the superinsulator.
  In the second type, as the magnetic duality, we introduce a duality between the ferromagnet and
   the superconductor. We applied the duality conditions by dual canonical formalism to the systems
    competing with each other, and we derived the relationship between the phase and the number of
     particles in a dual system of each other. We showed that we can more clearly define the presence
      of quantum resistance and a quantum critical point by a dual canonical formalism. In the second results,
       we indicate the conditions for a spin blockade in which electron tunneling does not take place,
        and as its application, we introduced a quantum spin transistor. 
\section*{Acknowledgments}
\begin{flushleft}
We acknowledge helpful discussions with S. Shinohara, K.Honma and A.Fujimoto.
\end{flushleft}


\end{document}